\documentclass[aps,prl,reprint,superscriptaddress, longbibliography]{revtex4-2}

\usepackage{lipsum, babel}
\usepackage{siunitx}
\usepackage{amsmath,bm}
\usepackage{amsbsy}
\usepackage{amssymb}
\usepackage{mathptmx, textcomp}
\usepackage{color}
\usepackage{braket}
\usepackage{graphicx}
\usepackage{textcomp}
\usepackage{notes2bib}
\usepackage{textcomp}
\usepackage{mwe}
\usepackage{siunitx}
\usepackage{filecontents}
\usepackage{soul}

\definecolor{bl}{rgb}{0, .1, .6}
\usepackage[colorlinks=true, citecolor = bl, linkcolor = bl, urlcolor=bl, pdfborder={0 0 0}]{hyperref}

\DeclareSIUnit\gauss{G}

\begin{document}
	
\title{Emergence of second-order coherence in the superradiant emission from a free-space atomic ensemble}
	
\author{Giovanni Ferioli}\email{ferioli@lens.unifi.it}
\affiliation{Dipartimento di Fisica e Astronomia, Universit`a degli studi di Firenze,
	Via G. Sansone 1, 50019 Sesto Fiorentino, Italy}	
\author{Igor Ferrier-Barbut}\email{igor.ferrier-barbut@institutoptique.fr}
\author{Antoine Browaeys}\email{antoine.browaeys@institutoptique.fr}
\affiliation{Universit\'e Paris-Saclay, Institut d'Optique Graduate School, CNRS, 
		Laboratoire Charles Fabry, 91127, Palaiseau, France}

\begin{abstract}
We investigate the evolution of the second-order temporal coherence during the emission of 
a superradiant burst by an elongated cloud of cold Rb atoms in free space. 
To do so, we measure the two-times intensity correlation function $g_N^{(2)}(t_1, t_2)$ following the pulsed excitation of the cloud. 
By monitoring $g_N^{(2)}(t, t)$ during the burst, we observe the establishment of second-order coherence, 
and contrast it with the situation where the cloud is initially prepared in a steady state.  
We compare our findings to the predictions of the Dicke model, using an effective atom number 
to account for finite size effects, finding that the model reproduces the observed trend at early time. 
For longer times, we observe a subradiant decay, 
a feature that goes beyond Dicke's model. 
Finally, we measure the $g_N^{(2)}(t_1, t_2)$ at different times and observe the appearance of anti-correlations
during the burst, that are not present when starting from a steady state. 
\end{abstract}
	
\date{\today}	
	
\maketitle

It is well known that a single excited two-level emitter 
decays exponentially, with a time constant related to the linewidth of the transition. 
This situation can be drastically modified if many emitters are coupled to a common radiation mode, 
as is the case when placed in an optical cavity or if they are confined within a subwavelength-sized volume. 
There, the system emits a short burst of photons with an initial increase of the emission rate.
The origin of this superradiant (or superfluorescent) burst, 
predicted by R. H. Dicke in 1954 \cite{dicke1954coherence,gross1982}, 
stems from a progressive phase locking  of the atomic dipoles. 
This happens spontaneously during the decay process, and results into  the $N^2$ scaling 
of the largest emission rate of the ensemble, a behavior considered a fingerprint of Dicke's superradiance.

Starting from the first experimental demonstrations in thermal gases of atoms and 
molecules \cite{PhysRevLett.30.309, PhysRevLett.36.1035, PhysRevLett.39.547}, 
superradiant bursts have been observed in many experimental platforms \cite{gross1982}, 
including nuclei \cite{rohlsberger2010collective}, artificial atoms \cite{mlynek2014observation} 
and recently a in cascaded atomic system \cite{PhysRevX.14.011020}. 
However, most of these works measured the photon emission rate  
while the investigation of other defining characteristics of the emitted light, such as first and second-order 
temporal and spatial coherence 
have been addressed only recently in few cases 
\cite{jahnke2016giant, gold2022spatial, tebbenjohanns2024predictingcorrelationssuperradiantemission, bach2024emergence}. 
Beyond its fundamental interest, exploring the statistical properties of a system 
featuring superradiant correlations will be useful to characterize a
novel class of superradiant lasers \cite{haake1993super, bohnet2012steady, debnath2018lasing, Maier:14} 
or other non-classical sources of light \cite{Tziperman:23, masson2020many} .

Here, we investigate the establishment of second-order temporal coherence during the emission of 
a superradiant burst \cite{PhysRevA.5.1457, tebbenjohanns2024predictingcorrelationssuperradiantemission} 
by an elongated cloud of laser cooled $^{87}$Rb atoms in free space. To do so, we 
measure the two-times intensity correlation function $g_N^{(2)}(t_1, t_2)$ of the emitted light during the decay
following the excitation of the cloud. 
Our results present strong similarities with a recent experiment carried out in an atomic ensemble 
chirally coupled to the mode of a nano-fiber \cite{bach2024emergence}. 
Here instead, the cooperative emission occurs symmetrically in two diffraction modes with opposite directions along the long axis 
of the cloud. 
Our findings confirm that the establishment of temporal correlations 
is a generic feature of Dicke superradiance, independent of the specific symmetries 
characterizing the atomic ensemble. We also find that, for a part of the emission, 
the experimental photon emission rate and equal-time correlation function 
are quantitatively reproduced by the Dicke model, introducing 
an effective atom number to account for the finite size extension of the cloud \cite{gross1982, ferioli2023non}. 
However, some observed features are, as expected, beyond the Dicke model, which fails for instance
to reproduce the observed subradiant decay at long time.  
This calls for a more sophisticated theoretical model 
able to include finite size effects as well as photon-mediated interaction between the emitters 
\cite{tebbenjohanns2024predictingcorrelationssuperradiantemission, mink2023collective,
bigorda2023characterizing, bigorda2022superradiance, francis2021theoretical}.

\begin{figure*}[!]
	\includegraphics[width=\linewidth]{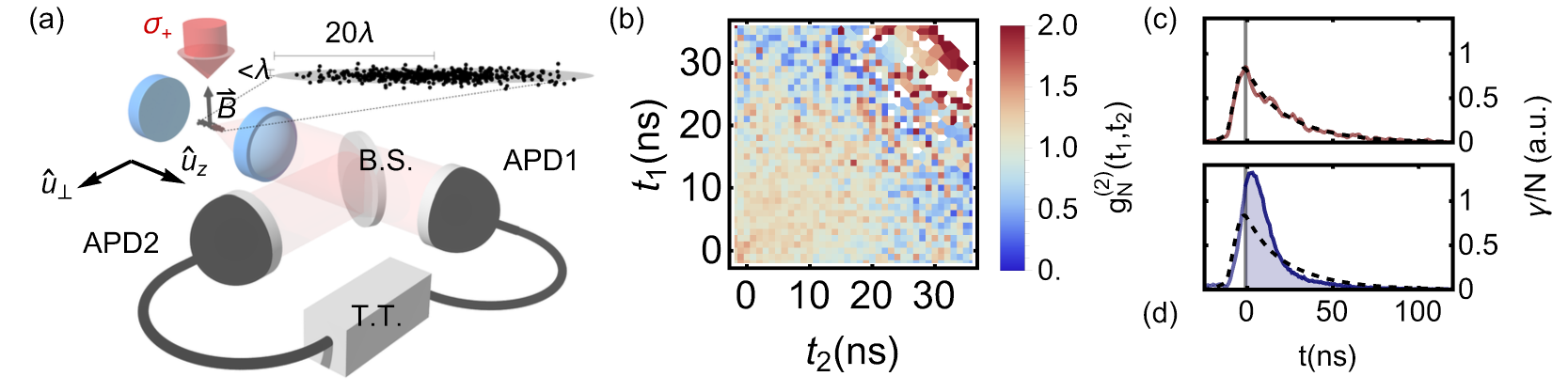}
	\caption{  
	(a) \textbf{ Experimental setup:} an elongated cloud of $^{87}$Rb atoms, 
	prepared in a dipole trap (not shown)
	is excited using a resonant laser beam propagating along a radial direction of the cloud.  
	The light emitted in the diffraction pattern, along ${\bf u}_z$, is split by a 50/50 beamsplitter (BS) 
	and sent onto two fibered avalanche photodiodes APD1,2. 
	A time tagger (T.T.) records the photon arrivals times on each photodiode
	with respect to a common trigger. 
	(b) Measurement of the two-times correlation function during the superradiant burst. 
	(white part: absence of any measured coincidences)  
	The number or repetitions is 30400, corresponding to about 10 hours of acquisition.
	(c,d) Photon emission rate collected in the ${\bf u}_z$ direction for $N\simeq 300$(c) 
	and $N\simeq 3000$(d),together with the solution 
	of the Optical Bloch Equations (dashed black line). 
	The vertical gray line indicates the end of the excitation pulse and 
	the color-filled region in (d) indicates the temporal window where the intensity correlation is measured. }
	\label{fig1}
\end{figure*}

Our experimental setup is sketched in Fig.\,\ref{fig1}(a)~\cite{glicenstein2021preparation}. 
We load up to $\simeq 3000$ $^{87}$Rb atoms in a $\SI{2.5}{\micro\meter}$-waist 
optical trap using gray molasses. 
The resulting elongated atomic cloud has a typical temperature of $\SI{200}{\micro \kelvin}$, 
a calculated radial r.m.s. size 
$\ell_{\text{rad}}\approx 0.6 \lambda$ and a measured axial r.m.s. size $\ell_{\text{ax}}\simeq 20 \lambda$ ($D_2$ transition, 
$\lambda= \SI{780.2}{\nano\meter}$, $\Gamma= 2\pi\times \SI{6}{\mega\hertz}$ 
and $I_{\text{sat}}\simeq  \SI{1.67}{\milli\watt/\centi\meter^2} $). 
To isolate two internal states and obtain a cloud of two-level atoms, 
we apply a $\SI{50}{\gauss}$-magnetic field oriented along a {\it radial} direction of the atomic cloud.  
The atoms are initially optically pumped to $\ket{g}=\ket{5S_{1/2}, F=2, m=2}$.  
To initiate a superradiant burst, we apply a pulse of resonant $\sigma^+$-polarized light
to excite the atoms in $\ket{e}=\ket{5P_{3/2}, F=3, m=3}$. 
The excitation beam is sent along the magnetic field, {\it perpendicularly} to the elongated axis of the cloud. 
Its duration is  \SI{12}{\nano\second} for a  Rabi frequency 
$\Omega \simeq 6.5 \Gamma$, limited by the available laser power. 
The excitation beam waist ($\SI{100}{\micro\meter}$) is much larger than the cloud dimensions, and
the optical density of the cloud along the beam direction is negligible so that  
all atoms experience the same $\Omega$.
The excitation pulse is temporally shaped with a combination of acousto- 
and electro-optical modulators, resulting in a turning on and off time of $\simeq\SI{1}{\nano\second}$. 
The photons emitted by the cloud along its main axis ($\bf u_z$ in Fig.\,\ref{fig1}(a)) 
are collected using fibered-coupled avalanches photodiodes (APDs). 

The excitation is performed with the trapping laser turned off to eliminate the inhomogeneous broadening it induces. 
To accumulate enough statistics, the same atomic cloud is excited up to 100 times 
before a new ensemble is prepared and the whole sequence repeated a few hundred times. 
The fraction of atoms lost during the excitation does not exceed $10\%$.
Given the temperature of the cloud, the atomic displacement $\delta r$ during 
the excitation and the emission by the cloud is negligible ({\it i.e.} $\delta r\ll 1/k$). 
We use an  Hanbury-Brown and Twiss (HBT) configuration to measure 
the intensity correlation function $g_N^{(2)}(t_1, t_2)$: 
we split the collected light in two via a 50/50 fiber beamsplitter. 
We then measure in each arm the photon arrival time with respect to a common trigger. 
From these, we compute $g_N^{(2)}(t_1, t_2)=  n_c(t_1,t_2)/[n_1(t_1)n_2(t_2)]$ 
where $n_i(t_k)$ is the photon number detected in arm $i$ at time
$t_k$ and $n_c(t_1,t_2)$ the number of coincidences on both arms at
times $t_1$ and $t_2$, as detailed in \cite{ferioli2024nongaussian}. 
The minimum time bin we use is 1 ns. We  present in Fig.\,\ref{fig1}(c) 
an example of two-times correlation function measured during a superradiant burst. 

Figure \,\ref{fig1} shows the photon emission rate $\gamma(t)$ measured on APD1
for a cloud containing $N\simeq300$ (c) and $N\simeq3000$ (d) during and after the excitation pulse. 
The data are compared to the solution of the Optical Bloch Equations  for a two-level atom (OBEs), 
including the measured temporal profile of the excitation laser. 
For low $N$, they are well reproduced by OBEs, indicating that the atoms
are uncorrelated. In contrast, we observe a superradiant burst at large $N$, 
as we showed in \cite{ferioli2021laser}.
For the Rabi frequency used and the finite temporal duration of the pulse, 
we reach a population inversion of about $85\%$, independent of $N$
\cite{ferioli2021laser}. We stress that the cloud is excited perpendicularly to its long axis, 
so that, even for a non-perfect $\pi$-pulse, the laser does not impart 
any phase coherence between the dipoles that could lead to an emission in the superradiant direction. 

\begin{figure}[!] 
	\includegraphics[width=\linewidth]{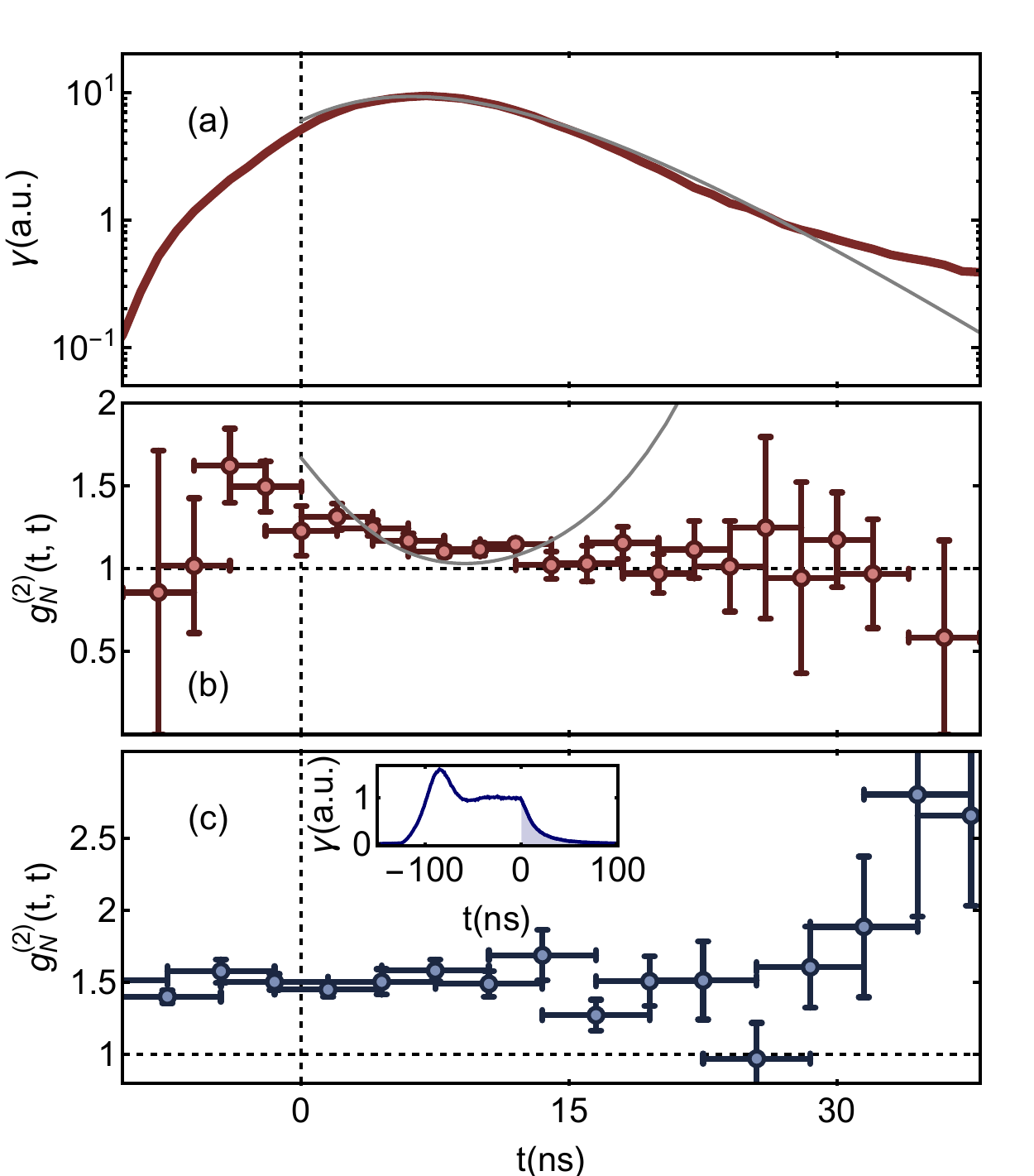}
	\caption{
	\textbf{Statistics of the light emitted during a superradiant burst:} 
	(a) red line: photon emission rate $\gamma(t)$, gray line prediction of DM (see text). 
	(b) Red points: measured values of $g_N^{(2)}(t,t)$ compared to the DM (gray line).
	Black dashed lines: guide to the eyes, indicating the end of the driving light ($t=0$) and $g_N^{(2)}(t,t)=1$, in (b) and (c). 
	Error bars in (b): for the y axes they correspond to the standard errors on the mean; 
	for x-axes they account for the number of bins used in the averaging process. 
	(c) Measurement of $g_N^{(2)}(t,t)$ after the system reaches the steady-state. 
	Error bars as in (b). 
	Inset: temporal traces showing the coherent dynamics 
	towards the steady state and the decay process 
	(dashed region) where $g_N^{(2)}(t,t)$ is measured.
	}
	\label{fig2}
\end{figure}

The main experimental result of this work is shown in Fig.\,\ref{fig2}, 
where we compare the photon emission rate and 
the equal time intensity correlation $g_N^{(2)}(t,t)$. 
This quantity is extracted from the two-times correlation function $g_N^{(2)}(t_1, t_2)$, 
considering $t=t_1=t_2$, and integrating over a time-bin of 2\,ns. 
We observe that the equal-time correlations decrease from $\sim 1.5$ 
right after switching off the excitation light down to $\sim 1$ after $\simeq 15\,{\rm ns}\simeq 1/2\Gamma$. 
This behavior signs the spontaneous establishment of second-order coherence due to superradiance: 
at the begin of the superradiant emission, the emitters are still nearly independent, 
leading to $g_N^{(2)}(t,t) >1$. Later, the progressive phasing of the dipoles during the superradiant emission 
results into an increasing second-order coherence, until $g_N^{(2)}(t,t) \approx 1$. 
Interestingly, $g_N^{(2)}(t,t)$ reached up to 1.7 for $t<0$ and start to decay during the pulse itself: 
this is consistent with our previous finding\,\cite{glicenstein2022superradiance} that superradiance 
starts to phase lock the dipoles even during the excitation. 
Finally, after a time $\sim 2/\Gamma$, the emission  becomes subradiant, 
while the correlations remain close to 1. 

In order to prove that the observed behavior is characteristic of the superradiant burst, 
we contrast it with the one where the atomic ensemble has reached steady-state under 
a constant drive, therefore preparing a different initial state. 
To do so, we apply the excitation laser with $\Omega =2\Gamma$
for a duration $\simeq \SI{120}{\nano\second}$, long enough for the system to reach steady-state. 
We then measure $g_N^{(2)}(t, t)$ after switching off of the driving laser, during the de-excitation process. 
From previous works, we know that the collective emission does not exhibit 
a burst but is rather 
characterized by a monotonous decay featuring mainly two time-scales 
\cite{ferioli2020storage, glicenstein2022superradiance}: a first, superradiant one, 
faster than the natural decay time ($\simeq \SI{12}{\nano\second}$ in this specific case, 
compared to $1/\Gamma\simeq 26$\,ns) 
and a slower, subradiant, one ($\simeq \SI{47}{\nano\second}$). 
\footnote{For the specific parameters chosen here, 
a few percent of the total excitation are stored in the subradiant modes.}. 
As we showed in Ref.\,\cite{ferioli2024nongaussian} 
(see also \cite{lassegues2022field, lassegues2024transition}), 
atomic correlations are established during the drive and result into a 
\emph{steady-state} value of the correlation $g_{N,\rm st}^{(2)}(0)$ lower than 2, 
the value expected for uncorrelated emitters. 
Figure \ref{fig2}(c) presents the result of the experiment: we observe that starting 
from $g_N^{(2)}(0, 0)= g_{N,\rm st}^{(2)}(0)$, the equal time intensity correlation 
remains constant until $\sim 30$\,ns, after which it increases.  
This trend deviates qualitatively and quantitatively from the one measured during 
the superradiant burst, indicating that the establishment 
of second-order temporal coherence is indeed related to the initial full inversion of the system 
leading to the burst. 

As a guide to understand the experimental results, we compare them
to the predictions of the Dicke model (DM) \cite{gross1982}. 
Briefly, the model considers $N$ two-level emitters ($\ket{e}$ and $\ket{g}$) 
identically coupled to a single electromagnetic mode. 
This indistinguishably allows treating the system as a collective spin 
$\hat{S}^+ = \sum_{i=1}^N \hat{\sigma}_i^+ $, 
where $\hat{\sigma}_i^+ = \ket{e}\bra{g}_i$ is the single atom rising operator. 
If the initial state is symmetric with respect to the exchange of two atoms, 
as the fully excited one, 
this permutational invariance is preserved during the dynamics: 
the system is constraint to evolve in the symmetric subspace of dimension $N+1$.
In the absence of an external driving, 
the collective emission is then governed by the following Lindblad equation:
\begin{equation}\label{Eq:LindbladDM}
	\frac{d \hat{\rho}(t)}{dt} = \frac{\Gamma}{2} \Big{(}2\hat{S}^- \hat{\rho}(t)\hat{S}^+ - \hat{S}^+ \hat{S}^- \hat{\rho}(t) -  \hat{\rho}(t)\hat{S}^+ \hat{S}^-  \Big{)}
\end{equation}
which can be easily solved for $\hat{\rho}(t)$. 
The photon emission rate is 
$\gamma(t) = \langle\hat{S}^+\hat{S}^-\rangle= {\rm Tr} [\hat{S}^+\hat{S}^-\hat{\rho}(t)]$. 
The equal time intensity correlation function is 
$g_N^{(2)}(t,t) = \langle\hat{S}^+\hat{S}^+\hat{S}^-\hat{S}^-\rangle_t /\langle\hat{S}^+\hat{S}^-\rangle_t^2 
= {\rm Tr}[\hat{S}^+\hat{S}^+\hat{S}^-\hat{S}^-\hat{\rho}(t)]/{\rm Tr}[\hat{S}^+\hat{S}^-\hat{\rho}(t)]^2$.
The DM assumes that all the atoms occupy a sub-wavelength volume, 
a condition far from being fulfilled experimentally. 
As shown in Ref.\,\cite{gross1982}, in a mean-field approximation, 
the average dynamics of atomic observables is still described by
Eq.(\ref{Eq:LindbladDM}) for an elongated sample by
introducing an effective atom number $\tilde{N} =  \mu N$, where the parameter 
$\mu$ accounts for the coupling between the cloud and its diffraction mode 
\cite{ferioli2020storage,tebbenjohanns2024predictingcorrelationssuperradiantemission}. 
It depends only on the geometry of the cloud and in our case $\mu \simeq 0.002$, 
meaning that around $\tilde{N}=6$ atoms are cooperatively coupled to the diffraction mode. 
We report the prediction of the DM  for $\gamma(t)$ and $g_N^{(2)}(t,t)$ as gray lines
in Fig.\,\ref{fig2}(a,b).  
They describe the experimental trend during the first part of the dynamics. 
including the time-scale over which second-order coherence is established. 
We have not attempted to compare the data in Fig.\,\ref{fig2}(c) to the DM: 
although it should describe the decay, it is not valid during the laser excitation $perpendicular$
to the main axis of the cloud, thus preventing us from calculating the atomic state in steady-state
and the dynamics after the excitation is turned off.  

Not surprisingly, however, we also observe discrepancies with respect to the predictions of the DM. 
First, the Dicke model predicts
$g_{N, \rm DM}^{(2)}(0,0)=2-2/\tilde N\approx 1.6$ \cite{masson2020many}. 
This value is obtained considering a fully inverted initial state $\ket{e}^{\otimes N}=\ket{N/2,N/2}$. 
Experimentally, we measure a smaller value,  
which might be explained by the fact that the initial state 
exhibits partial coherence due to the establishment of superradiant correlations
in a timescale faster than the $\pi$-pulse \cite{ferioli2021laser}. 
Second,  we observe in Fig.\,\ref{fig2}(a) the appearance of subradiant decay 
at long-time, absent from the DM: 
As we found in previous works \cite{glicenstein2022superradiance, ferioli2020storage}, 
subradiant states are populated 
during the superradiant burst and during the $\pi$ pulse.  
When subradiance dominates ($t\gtrsim 1/\Gamma$),  
we observe that $g_N^{(2)}(t,t) \approx 1$ while instead the DM predicts 
bunched correlations \cite{bhatti2015superbunching, jahnke2016giant, highly2018gulfam}: 
the second-order  temporal coherence of our extended cloud 
is longer than the one predicted by Dicke's model. 
This failure of the DM is  expected: the finite cloud size and light-induced dipole interactions 
break the permutational symmetry of the DM, leading to a finite coupling between bright and dark (subradiant) 
states \cite{cipris2021subradiance}. 

\begin{figure}[!] 
	\includegraphics[width=\linewidth]{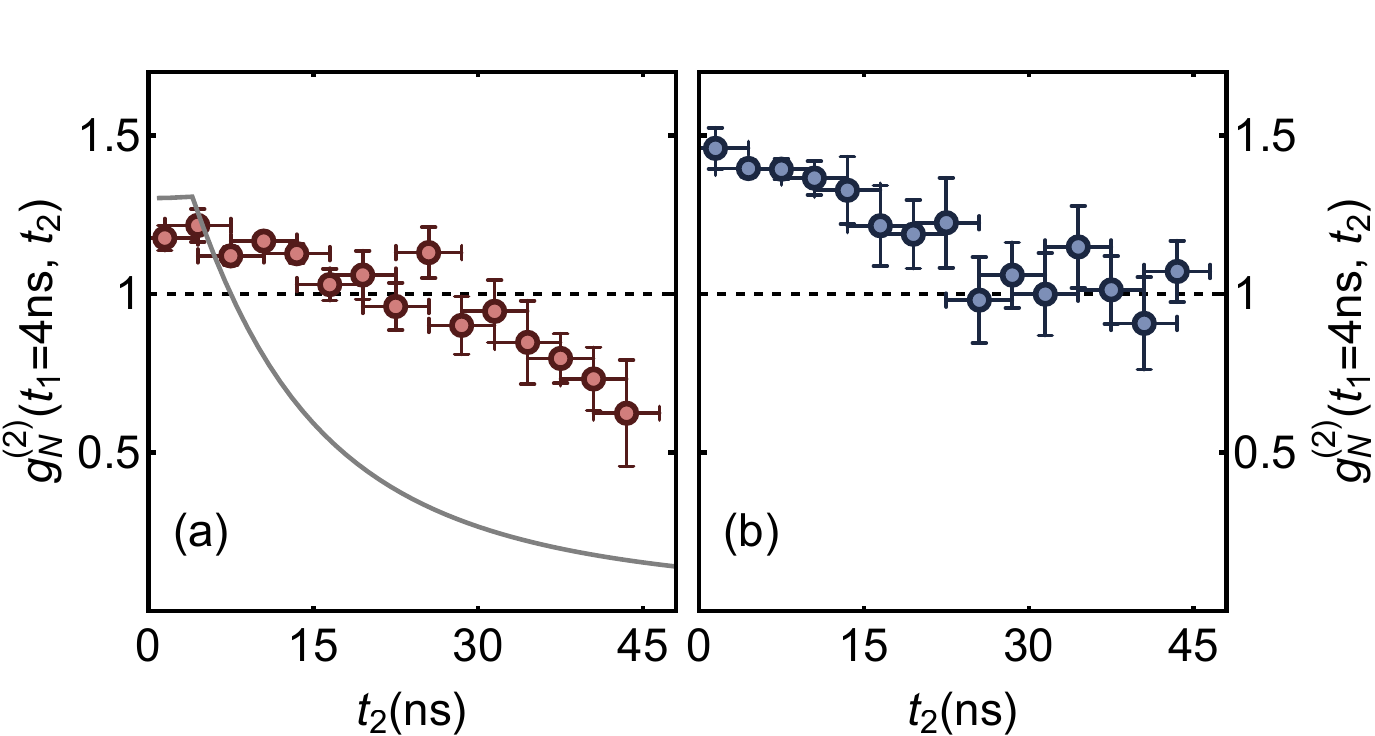}
	\caption{(a) Intensity correlations $g^{(2)}(t_1,t_2)$ at different times during a superradiant pulse,
		corresponding to a cut of the two-dimensional map shown in Fig.\,\ref{fig1}(b) for $t_1=4$\,ns (right). 
		The gray lines is the predictions of the Dicke Model for $\tilde N=6$, using the quantum
		regression theorem.
		(b) Same quantity measured during the decay following the steady-state 
		excitation of the cloud, as in Fig.\,\ref{fig2}(c). }
	\label{fig3}
\end{figure}

Finally, and along the lines explored in Ref.~\cite{bach2024emergence}, we plot in Fig.\,\ref{fig3} 
the two-times correlation functions $g^{(2)}(t_1,t_2)$ fixing $t_1=4$\,ns
and varying the second time $t_2$ (the observations below hold for different choices of $t_1$). 
As found in Ref.~\cite{bach2024emergence}, 
we observe here also the appearance of anti-correlations after 
$|t_1-t_2|\sim 25\,{\rm ns}\sim1/\Gamma$.  
As noted in Ref.~\cite{bach2024emergence}, one can interpret these anti-correlations 
as stemming from the fact that the superradiant burst emitted at each realization, 
triggered by a spontaneous emission at a random time, is shorter than the average 
intensity shown in Fig.\,\ref{fig1}(d) \cite{gross1982}. 
More generally, as the photon number $\sum_t n_1(t) = \sum_t n_2(t) = N_{\rm ph}$ 
emitted during the pulse following full inversion is fixed,
the observation of bunching implies that anti-correlations should be observed: 
Indeed, $\sum_{t,t'}n_c(t,t')=N_{\rm ph}(N_{\rm ph}-1)\simeq N_{\rm ph}^2$ 
and $\sum_{t,t'} n_1(t)n_2(t')=N_{\rm ph}^2$ lead to
$\sum_{t,t'}n_1(t)n_2(t')g^{(2)}(t,t')=\sum_{t,t'} n_1(t) n_2(t')$,
from which we conclude that $g^{(2)}(t,t)>1$ results into  $g^{(2)}(t,t')<1$ for some $t\neq t'$.
The quantitative predictions of the DM (using the quantum regression theorem \cite{steck2007quantum}) 
are also reported in Fig.\,\ref{fig3}(a): they reproduce the general trend, 
but fail to give the timescale, calling for a better model. 
Finally, we plot  $g^{(2)}(t_1,t_2)$ during the decay following the excitation of the cloud into a steady-state
(Fig.\,\ref{fig2}c). There, we do not observe the appearance of clear anti-correlations: the argument presented above
is indeed only valid when $N_{\rm ph}$ is the same for each repetition of the experiment. 
This is the case for a full inversion of the system leading to a superradiant burst, 
but not when preparing an atomic steady state, which is a mixed superposition of super and sub-radiant states  
containing various numbers of excited atoms: there, the number of emitted photons varies from shot-to-shot. 

In conclusion, we have measured the two-times intensity correlation function during a superradiant burst
emitted by an elongated atomic cloud, 
and  observed the establishment of second-order temporal coherence. 
The early dynamics of the correlations is reproduced by the Dicke model
introducing an effective atom number to account for the finite size. 
We have also observed the limit of the model, which
does not capture the subradiant decay at long-time. 
This poses the theoretical challenge of a microscopic description of the system 
\cite{agarwal2024directional, goncalves2024driven, mink2023collective}. 
Our work suggests the investigation of the 
intensity correlations also in the {\it driven} Dicke model \cite{ferioli2023non},  
where they are expected to oscillate at twice the Rabi frequency deeply in the 
superradiant phase \cite{agarwal1979intensity, hassan1982non}, 
a hallmark of collective behavior. 
Another direction is related to the study of directional effects in the photon statistics, 
which might reveal the many-body nature of the emission \cite{masson2020many, cardenaslopez2023manybody}, 
or the investigation of higher-order correlation functions \cite{stiesdal2018observation}

\begin{acknowledgments}  
We acknowledge discussions with M. Fleischhauer, C. Mink,  P. Schneeweiss, A. Rauschenbeutel and their team. 
We thank A. Glicenstein for assistance in the early stages of the experiment. 
This project has received funding from the European Research Council 
(Advanced grant No. 101018511, ATARAXIA), 
by the Agence National de la Recherche (ANR, project DEAR and ANR-22-PETQ-0004 France 2030, project QuBitAF). 
\end{acknowledgments}

\bibliography{g2_burst_biblio}

\end{document}